\documentclass[]{aastex631}

\usepackage[utf8]{inputenc}

\definecolor{armygreen}{rgb}{0.29, 0.33, 0.13}
\definecolor{darkpastelgreen}{rgb}{0.01, 0.75, 0.24}

\begin{document}

\title{Maps of solar wind plasma precipitation onto Mercury's surface: a geographical perspective}

\author{Federico Lavorenti}
\affiliation{Laboratoire Lagrange, Observatoire de la Côte d’Azur, Université Côte d’Azur, CNRS, Nice, France}
\affiliation{Dipartimento di Fisica ``E. Fermi'', Università di Pisa, Pisa, Italy}

\author{Elizabeth A. Jensen}
\affiliation{Planetary Science Institute, Tucson, AZ, USA}

\author{Sae Aizawa}
\affiliation{Dipartimento di Fisica ``E. Fermi'', Università di Pisa, Pisa, Italy}
\affiliation{Japan Aerospace Exploration Agency, Sagamihara, Kanagawa, Japan}

\author{Francesco Califano}
\affiliation{Dipartimento di Fisica ``E. Fermi'', Università di Pisa, Pisa, Italy}

\author{Mario D'Amore}
\affiliation{Institut für Planetenforschung, Deutsches Zentrum für Luft- und Raumfahrt (DLR), Rutherfordstr.
2, 12489 Berlin, Germany}

\author{Deborah Domingue}
\affiliation{Planetary Science Institute, Tucson, AZ, USA}

\author{Pierre Henri}
\affiliation{Laboratoire Lagrange, Observatoire de la Côte d’Azur, Université Côte d’Azur, CNRS, Nice, France}
\affiliation{LPC2E, CNRS, Univ. d’Orléans, CNES, Orléans, France}

\author{Simon Lindsay}
\affiliation{School of Physics and Astronomy, The University of Leicester, Leicester, UK}

\author{Jim M. Raines}
\affiliation{Dept. of Climate and Space Sciences and Engineering, University of Michigan, Ann Arbor, Michigan USA.}

\author{Daniel Wolf Savin}
\affiliation{Columbia Astrophysics Laboratory, Columbia University, MC 5247, 550 West 120th Street, New York, NY 10027, USA}

\date{January 2023}

\begin{abstract}
Mercury is the closest planet to the Sun, possesses a weak intrinsic magnetic field and has only a very tenuous atmosphere (exosphere). These three conditions result in a direct coupling between the plasma emitted from the Sun (namely the solar wind) and Mercury's surface. The planet's magnetic field leads to a non-trivial pattern of plasma precipitation onto the surface, that is expected to contribute to the alteration of the regolith over geological time scales.
The goal of this work is to study the solar wind plasma precipitation onto the surface of Mercury from a geographical perspective, as opposed to the local-time-of-day approach of previous precipitation modeling studies.
We employ solar wind precipitation maps for protons and electrons from two fully-kinetic numerical simulations of Mercury's plasma environment. These maps are then integrated over two full Mercury orbits (176 Earth days).
We found that the plasma precipitation pattern at the surface is most strongly affected by the upstream solar wind conditions, particularly by the interplanetary magnetic field direction, and less by Mercury's 3:2 spin-orbit resonance. We also found that Mercury's magnetic field is able to shield the surface from roughly 90\% of the incoming solar wind flux. At the surface, protons have a broad energy distribution from below 500 eV to more than 1.5 keV; while electrons are mostly found in the range 0.1-4 keV.
These results will help to better constrain space weathering and exosphere source processes at Mercury, as well as to interpret observations by the ongoing ESA/JAXA BepiColombo mission.
\end{abstract}

\section{Introduction}\label{sec:intro}
Mercury is the only telluric solar system planet, other than Earth, with an intrinsic magnetic field~\citep{Ogilvie1974}. Mercury's magnetic field shapes the interaction between the planet's surface and the surrounding solar wind --- a turbulent, supersonic, and magnetized plasma flowing outward from the Sun~\citep{Meyer-Vernet2007}. The interaction between Mercury's magnetic field and the solar wind is a crucial part of the global Hermean environment, both shape and evolution. This is a result of the planet's proximity to the Sun (0.31 AU at perihelion and 0.47 AU at aphelion) and its relatively weak magnetic field (around 200 nT at the surface).

The interaction between the solar wind and Mercury's magnetic field determines the precipitation of solar wind protons and electrons onto the planet's surface, through a complex series of coupled local processes. On the dayside, the presence of a bow shock in front of the planet along with magnetic reconnection in the magnetopause determines the pattern and energy of precipitating plasma; while on the nightside, the precipitating plasma is mostly affected by magnetic reconnection and the magnetic field configuration in the tail.
On the dayside, downstream of the bow shock (in the so-called magnetosheat), the plasma density, temperature and magnetic field increases while the plasma bulk velocity decreases. This slowdown leads to an order of magnitude reduction in the proton kinetic energy, which still dominates over the thermal energy component in the total energy of the protons. This differs with respect to the slowdown in the electron velocity, which does not alter significantly the electron total energy, as the thermal energy component dominates over the kinetic component in the total energy of the electrons. Downstream of the magnetosheat, at the magnetopause, a part of the magnetic field carried by the solar wind connects with Mercury's magnetic field through magnetic reconnection. This process allows a fraction of shocked solar wind plasma to precipitate onto the surface, spiraling along newly-opened magnetic field lines.
On the nightside, particles are accelerated and ejected planetward as magnetic field lines reconnect in the tail of the magnetosphere. This high-energy plasma is accelerated through the plasma sheet horns (i.e., regions connecting the plasma sheet to the surface at mid latitudes, analogous to auroral ovals at Earth) and precipitates onto the surface spiraling along magnetic field lines.
The magnetic topology of the planet 
determine, to a large extent, the geographical distribution of plasma precipitation on both sides of the planet.

\begin{figure*}[b]
\centering
    \includegraphics[width=\linewidth]{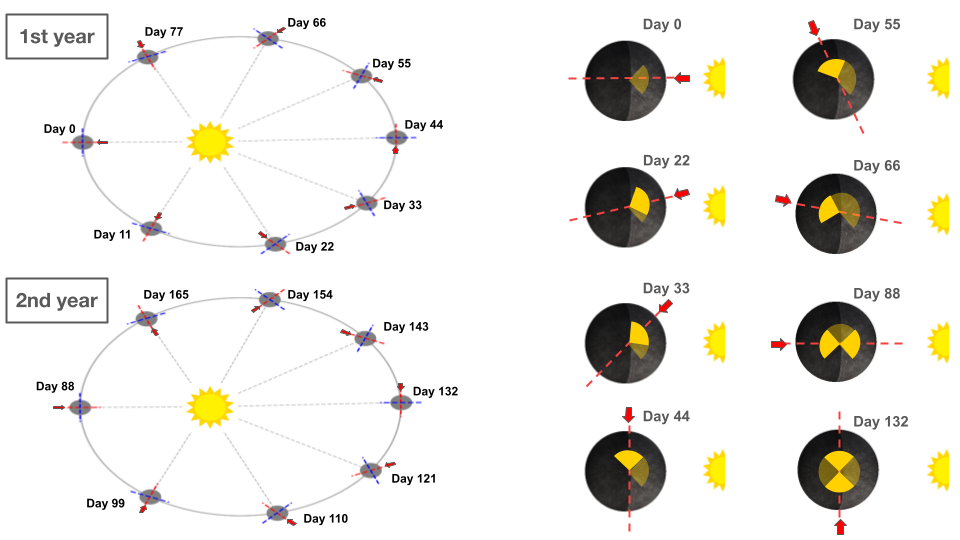}
    \caption{Sketch of plasma precipitation onto Mercury's surface over orbital time. The 3:2 spin-orbit resonance is responsible for increasing the cusp precipitation at the hot poles (red dashed lines) compared to the warm poles (blue dashed lines).}
    \label{fig:fig3_sketch}
\end{figure*}

Plasma precipitation onto the surface of Mercury is further affected by the 3:2 spin-orbit resonance of the planet, as shown in Fig.~\ref{fig:fig3_sketch}. As a consequence of this resonance, an observer standing at longitude $0^{\circ}$ (or $180^{\circ}$) faces the Sun for a longer time and at a closer distance compared to longitude $90^{\circ}$ (or $-90^{\circ}$). Therefore, Mercury's surface at longitudes $0^{\circ}$ and $180^{\circ}$ (called the hot poles) experience a higher mean photon flux than at longitudes $90^{\circ}$ and $-90^{\circ}$ (called the warm poles). This was confirmed by the NASA MESSENGER mission~\citep{Solomon2007}, which found a bimodal longitudinal pattern characterized by hot (warm) poles with maximal temperatures of 700$^{\circ}$K (570$^{\circ}$K)~\citep{Bauch2021}.
A similar longitudinal pattern is expected to arise in the plasma fluxes at the surface when integrating over two full Mercury orbits. Past numerical works have addressed this problem in terms of ``time of day'' (local time) on the surface~\citep{Benna2010,Schriver2011b,Fatemi2020,Lavorenti2023}, but plasma precipitation has yet to be examined with regard to geographic location. 

Looking at plasma precipitation from a geographic perspective will enable to correlate plasma fluxes with spectral and compositional properties of the surface and, thus, to explore potential causal relationships with space weathering driven by the solar wind. Precipitation of solar wind particles onto the surface of Mercury drives space-weathering processes such as ion sputtering, ion implantation, electron stimulated desorption (ESD), and X-ray fluorescence (XRF)~\citep{Domingue2014,Wurz2022}. Ion irradiation affects the surface at an atomic level and the exosphere at a global level. Ion sputtering is thought to be one of the main source processes for high altitude sodium in the Hermean exosphere~\citep{Mangano2015,Killen2022}. The maps presented in this work will allow researchers to better quantify ion sputtering at the surface of Mercury by relating the geographical distributions of ion fluxes with the surface distribution of sputtered species.
ESD is another poorly understood source process of Mercury's exosphere~\citep{Madey1998,McLain2011,Domingue2014}. The electron maps computed in this work, coupled with maps of surface temperature and composition, will enable a precise description of ESD for exosphere models.
XRF, driven by $\sim$~keV electrons, converts precipitating electrons to X-rays photons at the surface of Mercury~\citep{Lindsay2016,Lindsay2022,Lavorenti2023}. Future X-rays observations at Mercury by the joint ESA/JAXA BepiColombo mission~\citep{Benkhoff2021} will benefit from the electron-precipitation maps computed in this work. The BepiColombo/MIXS instrument~\citep{Bunce2020} will be able to constrain the surface composition and mineralogy, in part, by using solar wind electrons to ``probe'' the Hermean surface via XRF.

Here, we present the first plasma precipitation maps at Mercury integrated over two full Mercury orbits (176 Earth days) to account for the spin-orbit resonance. We use the proton and electron precipitation maps published in~\citet{Lavorenti2023} as a function of ``time of day'' as inputs to our computations over Mercury's orbit. In this work, we neglect heavy solar wind ions (with atomic number $Z \ge 2$), micro-meteoroids impacts, and thermal processes acting at the surface. The rest of this paper is organized as follows. Section~\ref{sec:methods} describes the methods used in this work. Section~\ref{sec:results} presents the results with a focus on the spatial and energy distribution of particles at the surface of Mercury. In Section~\ref{sec:discussion}, we discuss the implications of our results for Mercury science and, more broadly, for space weathering of weakly magnetized bodies.
Section~\ref{sec:conclusion} summarizes the conclusions of the paper.

\begin{figure}[b]
    \centering
    \includegraphics[width=0.6\linewidth]{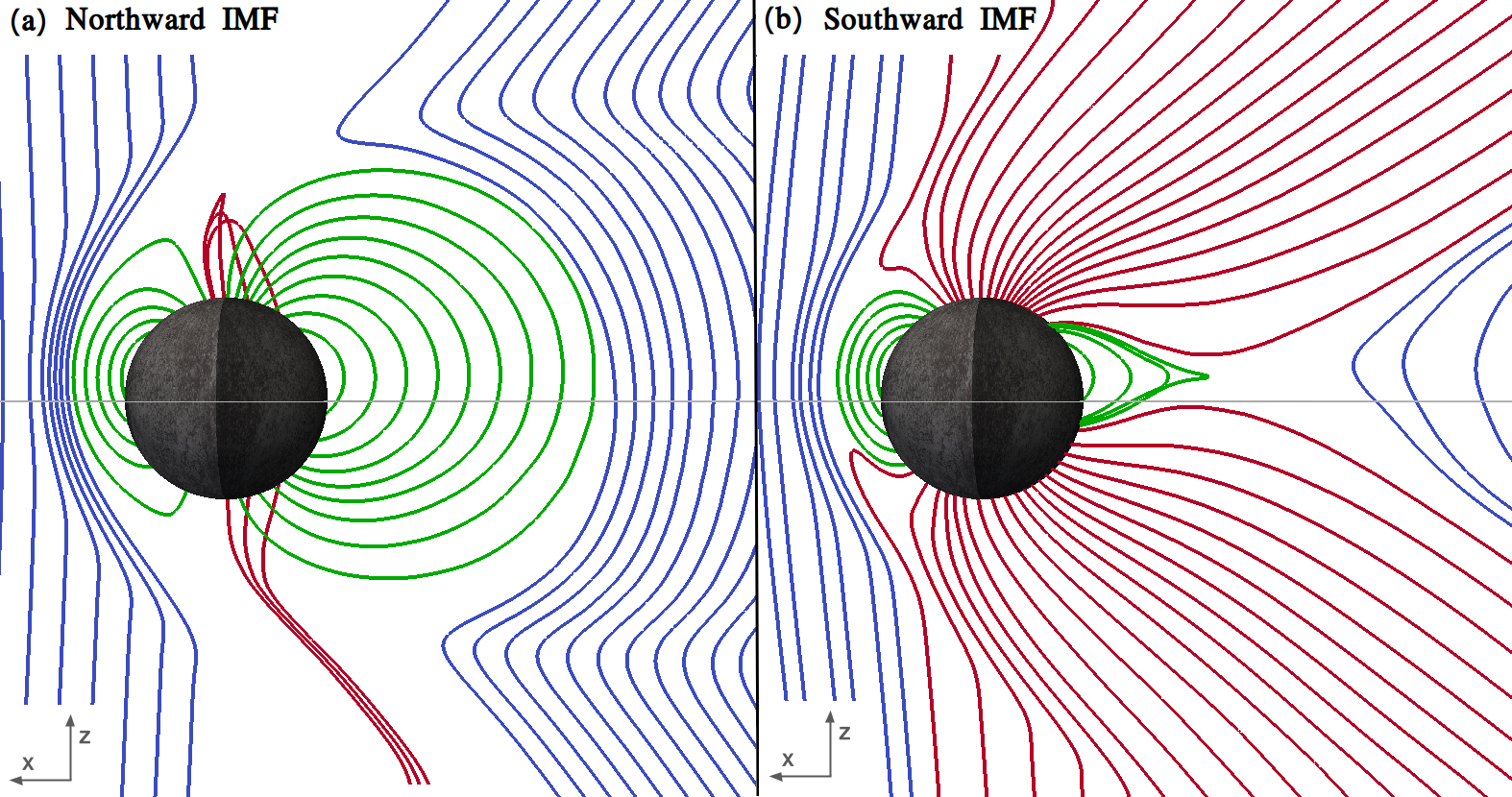}
    \caption{Magnetic field lines in the meridian plane (the X-Z plane in the Mercury-centered Solar Orbital (MSO) coordinate system) for our two simulations. The left (right) panel corresponds to a simulation with a northward (southward) IMF.
    Blue lines correspond to magnetic field lines not connected to the planet (i.e., solar wind field lines).
    Red lines correspond to magnetic field lines with one end connected to the planet (i.e., magnetospheric open field lines).
    Green lines correspond to magnetic field lines with both ends connected to the planet (i.e., magnetospheric closed field lines).
    The horizontal gray lines indicate the geographic equator of the planet.}
    \label{fig:Blines}
\end{figure}

\section{Methods}\label{sec:methods}
We utilize the proton and electron precipitation maps published in~\citet{Lavorenti2023}, which were computed using a fully-kinetic, global, three-dimensional plasma model of Mercury's magnetosphere. This numerical model solves the Vlasov-Maxwell system of equations using an implicit PIC (Particle-In-Cell) algorithm~\citep{Markidis2010}. The model solves the plasma dynamics of the interaction between the solar wind and Mercury's magnetic field, self-consistently including the kinetic physics of both protons and electrons. Kinetic models differ from other fluid plasma models (e.g., magnetohydrodynamic models) that do not take into account the velocity distribution functions of the particles but use only averaged quantities such as density, bulk velocity, pressure, etc. In our simulations, the normalized planetary radius, the proton-to-electron mass ratio, and the light-to-Alfv\'en speed ratio are artificially reduced in order to be able to run on state-of-the-art HPC (High Performance Computing) facilities, while maintaining a good – although compressed – separation of scales between the planetary radius, the proton gyroradius, and the electron gyroradius. In~\citet{Lavorenti2022}, the authors validated these compressed length scales by comparison of the model results with \textit{in situ} MESSENGER observations averaged over the mission time period.

The solution of our numerical model depends upon the upstream solar wind parameters. We use solar wind (SW) parameters corresponding to typical mean values at Mercury's aphelion~\citep{Sarantos2007,James2017} with a plasma density $n_{_{\text{SW}}}$ = 30 cm$^{-3}$, speed $V_{_{\text{SW}}}$ = 400~km~s$^{-1}$, magnetic field amplitude $B_{_{\text{SW}}}$ = 20 nT, and proton and electron temperatures $T_{{\text i,_{\text{SW}}}}$ = $T_{{\text e,_{\text{SW}}}}$ = 21.5 eV. These parameters are kept fixed in our two numerical simulations, while we vary the direction of the interplanetary magnetic field (IMF) in our two runs from purely northward to purely southward. Different directions of the IMF corresponds to different magnetic configurations at the surface of Mercury as shown in Fig.~\ref{fig:Blines}. These configurations in turn affects plasma precipitation at the surface. In this work, we address two extreme IMF configurations (with $B_x = B_y=0$) rarely found in the real system, but which are useful to grasp the role of the magnetic field in shaping the plasma precipitation at the surface. Our IMF configurations show the range of variability of the system, by providing conditions for minimal (maximal) magnetic coupling between the solar wind and the planet when the IMF is northward (southward), i.e., when the IMF is anti-parallel (parallel) to the planetary magnetic dipole moment. 

Plasma precipitation maps as a function of local time were integrated along two full Mercury orbits, from 2022 January 23 at 18:44:32 UTC to 2022 July 18 at 05:48:44 UTC, this corresponds to the time interval shown in Fig.~\ref{fig:fig3_sketch}.
We do not consider variations in the solar wind parameters along the orbit, which will be included in a future work including also a more realistic IMF configurations.
We subdivide the orbit into 515 steps of equal time $dt$ = 8.21 hours. At each timestep, we rotate the planet and map local time to surface longitude using Jet Propulsion Laboratory's NAIF (Navigation and Ancillary Information Facility) SPICE files and routines. The NAIF/SPICE files enable determining precisely the location of the Sun with respect to Mercury at each time of the orbit, and consist of ephemeris files for the orbit of Mercury and planetary body information for its rotation. The geographical registration of local time to longitude through this time period is provided with the ancillary files to this publication~\citep{Jensen2023_zenodo}. 

\section{Results}\label{sec:results}

\subsection{Spatial distribution of particle fluxes at the surface}\label{subs:spatial_distrib}

\begin{figure*}
    \centering
    \includegraphics[width=\linewidth]{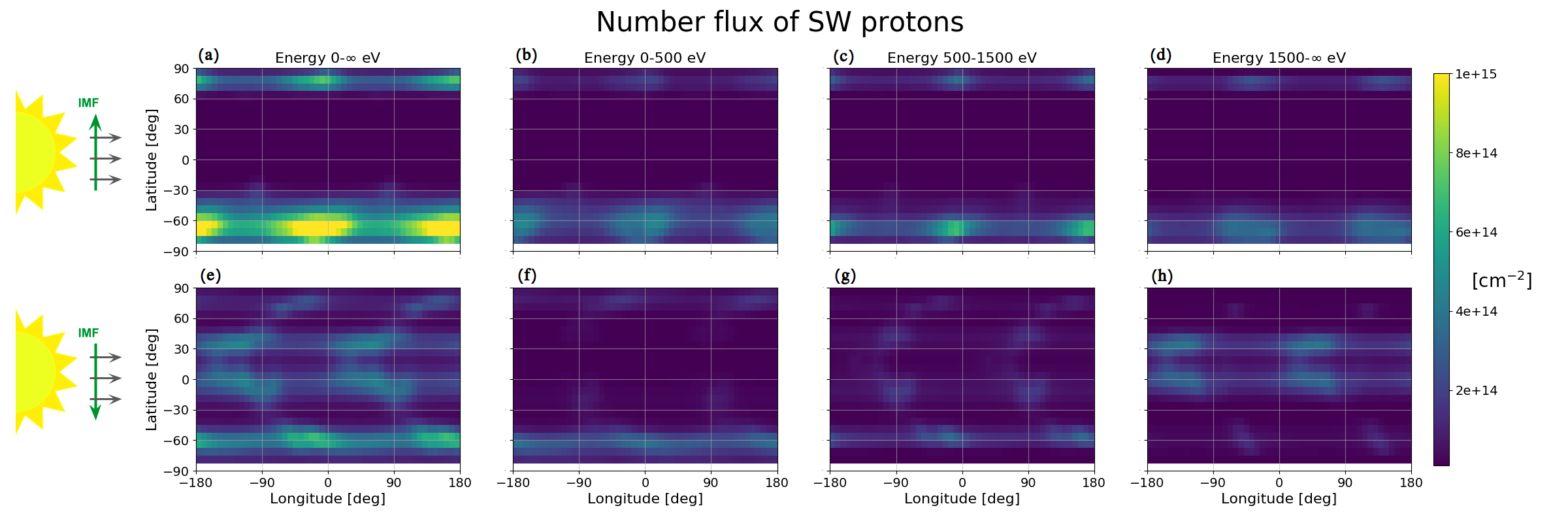}
    \caption{Proton precipitation maps integrated over two full Mercury orbits. Panels (a)-(d) are obtained from a simulation with a purely northward IMF and panels (e)-(h) with a purely southward IMF, respectively shown by the green vectors on the left. The different columns correspond to different energy bins, given at the top of each column. The grey horizontal arrows in the left sketches indicate the solar wind velocity. At times when longitude 0$^{\circ}$ is subsolar, longitudes -90$^{\circ}$ and +90$^{\circ}$ correspond to local dawn and dusk, respectively.}
    \label{fig:fig1_ions}
\end{figure*}

\begin{figure}
    \centering
    \includegraphics[width=0.49\linewidth]{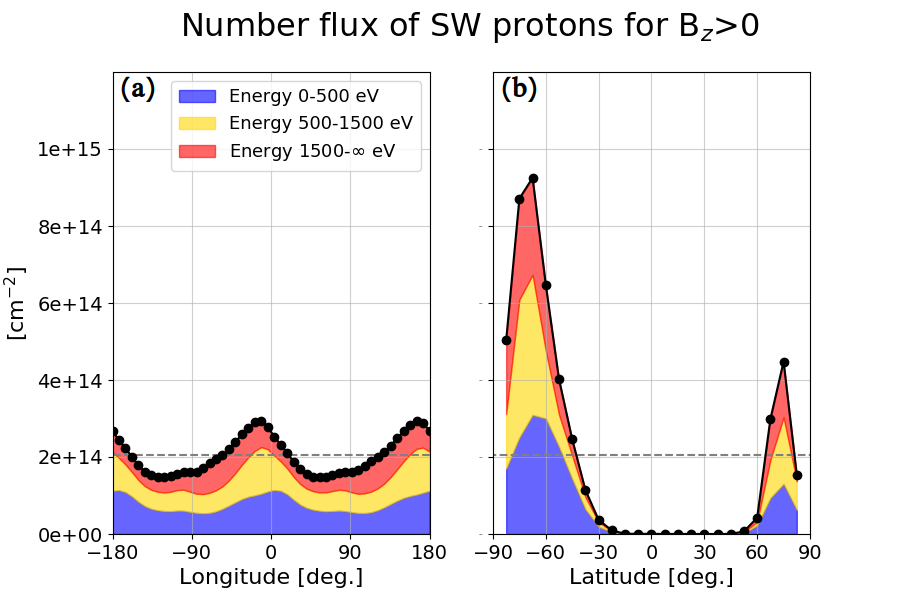}
    \includegraphics[width=0.49\linewidth]{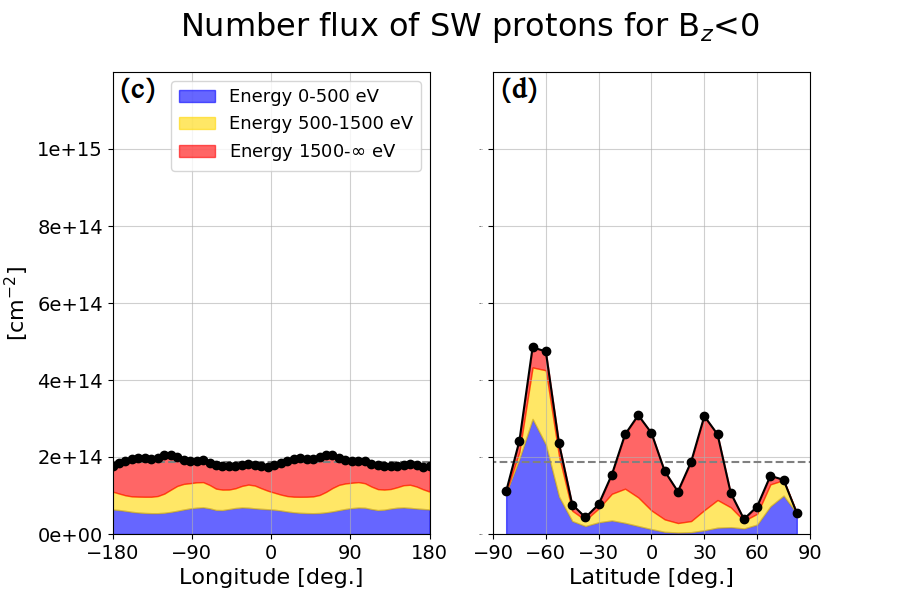}
    \caption{Proton precipitation maps averaged over latitude and longitude. The curves in panels (a) and (c) are obtained from the maps in Fig.~\ref{fig:fig1_ions}(a)-(d) and Fig.~\ref{fig:fig1_ions}(e)-(h), respectively, by averaging in latitude. The curves in panels (b) and (d) are obtained, respectively, from the same maps by averaging in longitude. The black dots and line show the averages summed over all energies and the colored areas the contributions for the labeled energy bins. The horizontal dashed black lines show the mean value of the fluence (averaged both in latitude and longitude).}
    \label{fig:fig2_means_i}
\end{figure}

\begin{figure*}
    \centering
    \includegraphics[width=\linewidth]{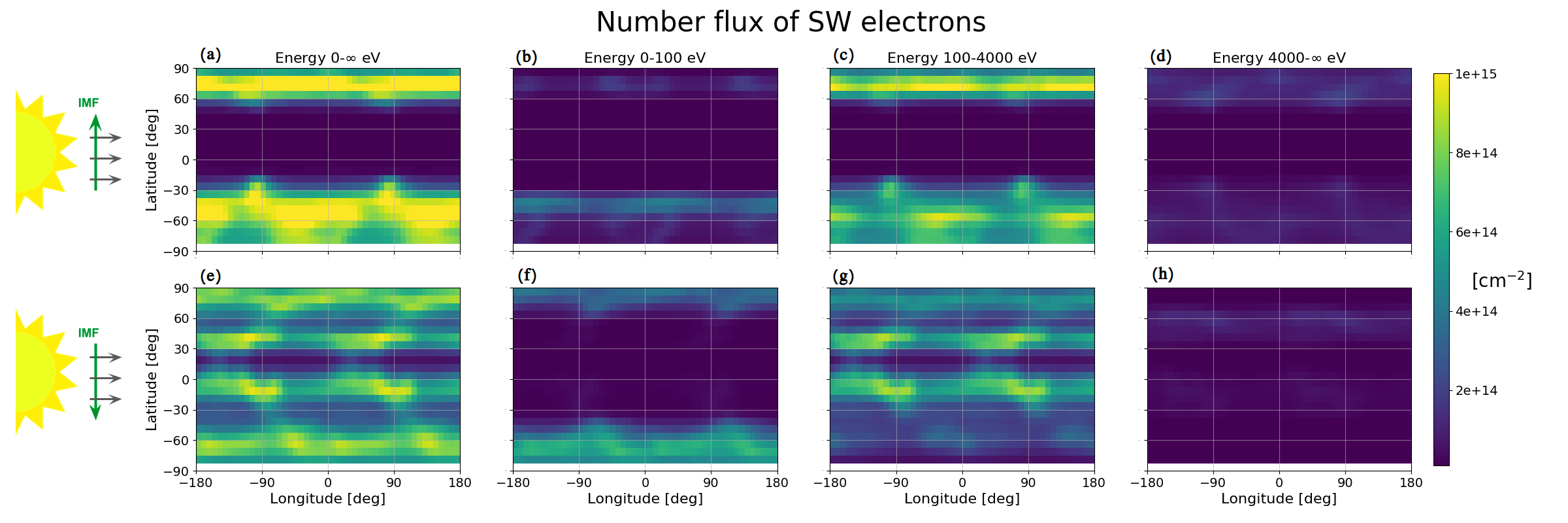}
    \caption{Same as Fig.~\ref{fig:fig1_ions} but for electron precipitation. }
    \label{fig:fig1_electrons}
\end{figure*}

\begin{figure}
    \centering
    \includegraphics[width=0.43\linewidth]{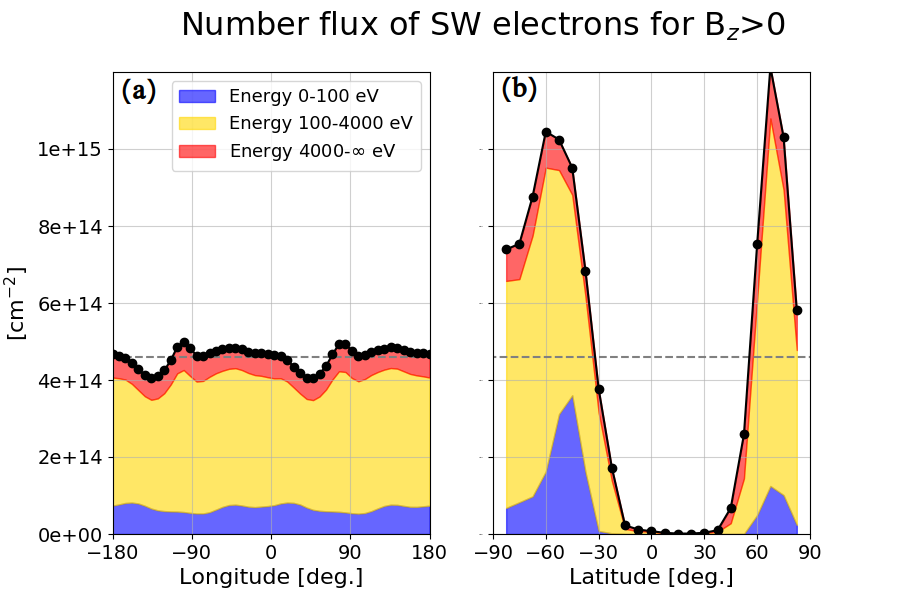}
    \includegraphics[width=0.43\linewidth]{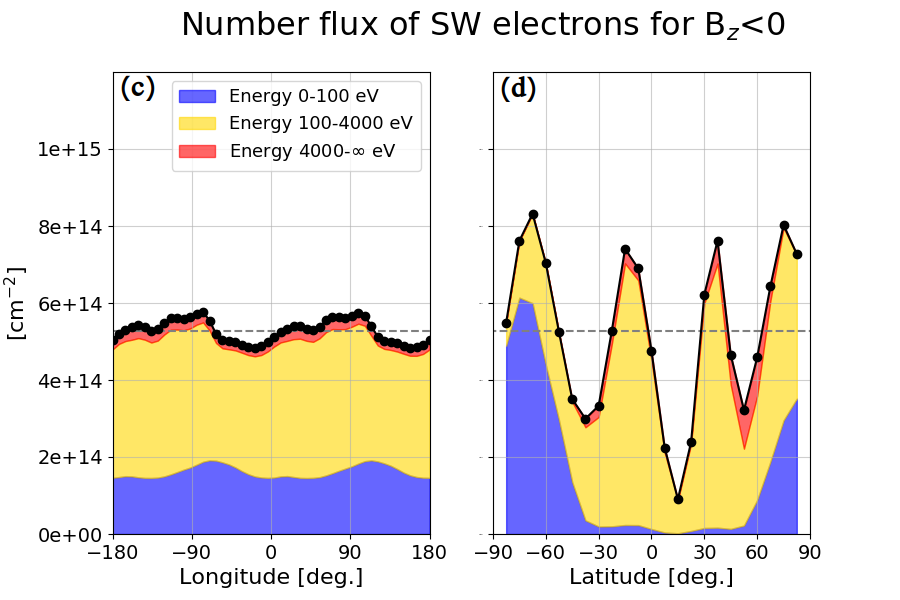}
    \caption{Same as Fig.~\ref{fig:fig2_means_i} but for electron precipitation.}
    \label{fig:fig2_means_e}
\end{figure}

The spatial distribution of particle fluxes at the surface is organized into latitudinal bands with enhancements in longitude driven by the 3:2 resonance. 
For protons, this pattern is shown by the maps in Fig.~\ref{fig:fig1_ions}, along with the latitudinal and longitudinal averages in Fig.~\ref{fig:fig2_means_i}.
The proton data are shown for energy bins of $0-500$, $500-1500$, and $1500-\infty$~eV, which we refer to as low, moderate, and high energies.
For electrons, Fig.~\ref{fig:fig1_electrons} and Fig.~\ref{fig:fig2_means_e} show the corresponding data. 
The electron data are shown for energy bins of $0-100$, $100-4000$, and $4000-\infty$~eV, which we refer to as low, moderate, and high energies.
The results from these maps are summarized in Tab.~\ref{tab:tab1_results_maps}, using a coarse spatial grid and averaging between the two IMF conditions under study.
In these maps, longitudinal variations are controlled by Mercury's rotation, while latitudinal variations are controlled by the IMF. In the following, we discuss these two effects separately.

Mercury's rotation is responsible for the differential accumulation of particles versus longitude. Due to the 3:2 spin-orbit resonance of Mercury, subsolar (local time 12 h) high-latitude proton precipitation is enhanced at the hot poles (longitude $0^{\circ}$ and $180^{\circ}$), as shown in Fig.~\ref{fig:fig1_ions}. This effect is more prominent in the simulation with a northward IMF, as shown in Fig.~\ref{fig:fig1_ions}(a)-(d) and Fig.~\ref{fig:fig2_means_i}(a). Under a northward IMF, the topology of the magnetosphere (i.e., the ``closed'' topology shown in Fig.~\ref{fig:Blines}(a)) channels plasma precipitation to the high-latitude cusps at local time 12 h. This hot pole enhancement drifts dawnward (by roughly -20$^{\circ}$ longitude, as shown in Fig.~\ref{fig:fig2_means_i}(a)) with increasing proton energy; we speculate that this drift is due to finite-Larmor-radius effects as the proton gyroradius increases with energy.
Under southward IMF and the resulting ``open'' magnetic field topology of the magnetosphere shown in Fig.~\ref{fig:Blines}(b), plasma precipitation occurs both at the cusps and at low latitudes. Cusp precipitation is enhanced at the hot poles due to the same mechanism at play with the northward IMF, as shown in Fig.~\ref{fig:fig1_ions}(e)-(g). This contrasts with the low-latitude proton precipitation, as shown in Fig.~\ref{fig:fig1_ions}(g)-(h). 
These low-latitude protons are ejected from the reconnection site in the tail and precipitate between roughly midnight and dawn in local time due to gradient-curvature drifts in the magnetosphere~\citep{Lavorenti2022,Glass2022}. Such proton precipitation around dawn in local time translates to precipitation around the warm poles in geographical coordinates.

Electrons precipitate onto the surface of Mercury with a pattern somewhat similar to that of the protons, as both are driven by the magnetic field topology. Longitudinal enhancements for electrons are weaker than for protons due to their higher thermal energy component (that accounts for random motion) as compared to their kinetic energy component (that accounts for ordered motion).
Under a northward IMF, electron precipitation in the cusps is enhanced around the hot poles at high latitude, as shown in Fig.~\ref{fig:fig1_electrons}(a)-(c). 
In this case, the ``closed'' magnetosphere topology channels electrons at high latitudes, and inhibits electron precipitation at low latitudes.
Under a southward IMF, the electron precipitation presents two distinct regions of precipitation, the high-latitude low-energy electrons in Fig.~\ref{fig:fig1_electrons}(f), and the low-latitude moderate-energy electrons in Fig.~\ref{fig:fig1_electrons}(g). Both tend to be enhanced around the warm poles, as shown in Fig.~\ref{fig:fig2_means_e}(c), although these enhancements are quite weak.
These enhancements at the warm poles are a direct consequence of electron precipitation toward dawn in local time~\citep{Lavorenti2023,Lindsay2022}.


Variations in the IMF drive strongly different latitudinal particle distributions at the surface.
Under a northward IMF, Mercury's magnetic dipole is able to shield a large fraction of the planet from the impinging solar wind.
In this case, protons and electrons precipitate mostly onto the cusps at latitudes spanning from $60^{\circ}$ to $80^{\circ}$ in the northern hemisphere and latitudes from $-30^{\circ}$ to $-70^{\circ}$ in the southern hemisphere, as shown in Fig.~\ref{fig:fig2_means_i}(b) for protons and Fig.~\ref{fig:fig2_means_e}(b) for electrons. The north-south asymmetry in the size of the cusp precipitation region is a consequence of the northward shift of Mercury's magnetic dipole. The position of the northern cusp is in agreement with recent estimates by~\citet{Raines2022} using MESSENGER/FIPS data. Under a northward IMF, particle precipitation is negligible in the equatorial region at latitudes spanning from $+60^{\circ}$ to $-30^{\circ}$.
Conversely, under a southward IMF, the topology of the magnetosphere results in stronger precipitation at low latitudes, as shown in Fig.~\ref{fig:fig2_means_i}(d) for protons and \ref{fig:fig2_means_e}(d) for electrons. This low-latitude precipitating plasma is composed of high-energy protons and moderate-energy electrons ejected planetward from the reconnection site in the tail. 
Respectively, comparing Fig.~\ref{fig:fig2_means_i}(b) and Fig.~\ref{fig:fig2_means_i}(d) with Fig.~\ref{fig:fig2_means_i}(a) and Fig.~\ref{fig:fig2_means_i}(c) for the protons, or equivalently Fig.~\ref{fig:fig2_means_e}(b) and Fig.~\ref{fig:fig2_means_e}(d) with Fig.~\ref{fig:fig2_means_e}(a) and Fig.~\ref{fig:fig2_means_e}(c) for electrons, we observe that latitudinal variations driven by the IMF are about one order of magnitude more important than longitudinal variations driven by Mercury's rotation. 

A coarse energy sampling of the precipitation maps in three energy bins, that are shown in Figs.~\ref{fig:fig1_ions}-\ref{fig:fig2_means_e}, enables one to identify the importance of various particle distributions in driving different processes at the surface.
The fluence of the particles in each of the three energy bins are summarized in Tab.~\ref{tab:tab1_results_maps}, using a coarse spatial grid and averaging over the two IMF configurations under study. In this table, we show the mean fluence of protons and electrons in each energy bin in three regions at the surface of Mercury, namely the north pole (above 60$^{\circ}$ latitude), the equatorial region (from 60$^{\circ}$ to $-30^{\circ}$ latitude) and the south pole (below $-30^{\circ}$ latitude). These fluence values can be used for a first order estimate of space-weathering due to plasma-surface interaction processes at Mercury. This point is further discussed in Sec.~\ref{sec:discussion}.


\begin{table*}
    \centering
    \begin{tabular}{|cc|c|c|c|}
         \hline
                                              & Energy [eV]     & North Pole & Equatorial region & South Pole \\
         \hline
         Proton fluence [10$^{14}$ cm$^{-2}$]   & 0-500         & 0.9        & small             & 1.8        \\
                                              & 500-1500      & 0.8        & small             & 1.2        \\
                                              & 1500-$\infty$ & 0.5        & 0.7               & 0.9        \\
         \hline
         Electron fluence [10$^{14}$ cm$^{-2}$] & 0-100         & 1.8        & small             & 2.8        \\
                                              & 100-4000      & 5.8        & 2.2               & 4.0        \\
                                              & 4000-$\infty$ & 0.7        & small             & small      \\
        \hline
    \end{tabular}
    \caption{Particle fluences in units of $10^{14}$~cm$^{-2}$ for different regions at the surface of Mercury, averaged over the two IMF simulations. North pole corresponds to latitudes from $+90{^{\circ}}$ to $+60{^{\circ}}$. Equatorial region corresponds to latitudes from $+60{^{\circ}}$ to $-30{^{\circ}}$. South pole corresponds to latitudes below $-30{^{\circ}}$. The entries ``small'' in the table indicate negligible fluence values that are below 0.5~$\times$~10$^{14}$~cm$^{-2}$.}
    \label{tab:tab1_results_maps}
\end{table*}

\begin{figure}
    \centering
    \includegraphics[width=0.49\linewidth]{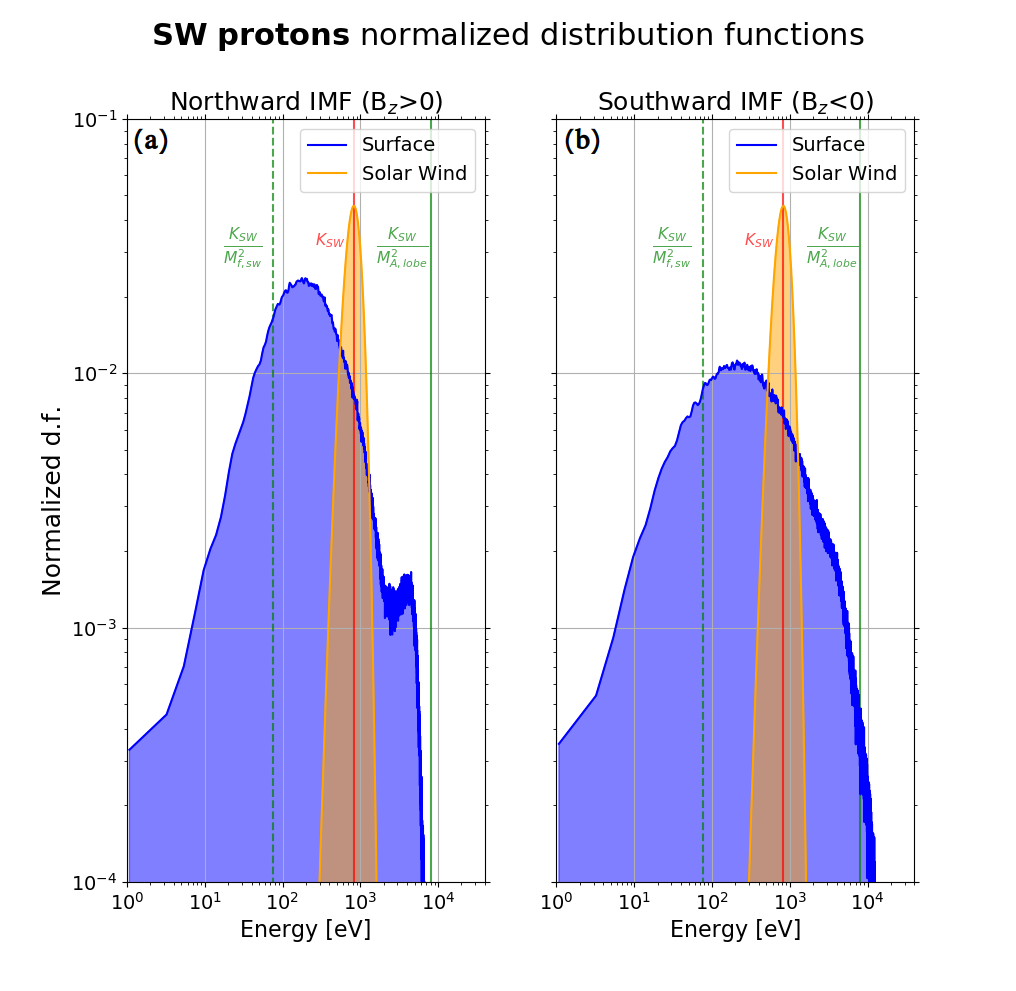}
    \includegraphics[width=0.49\linewidth]{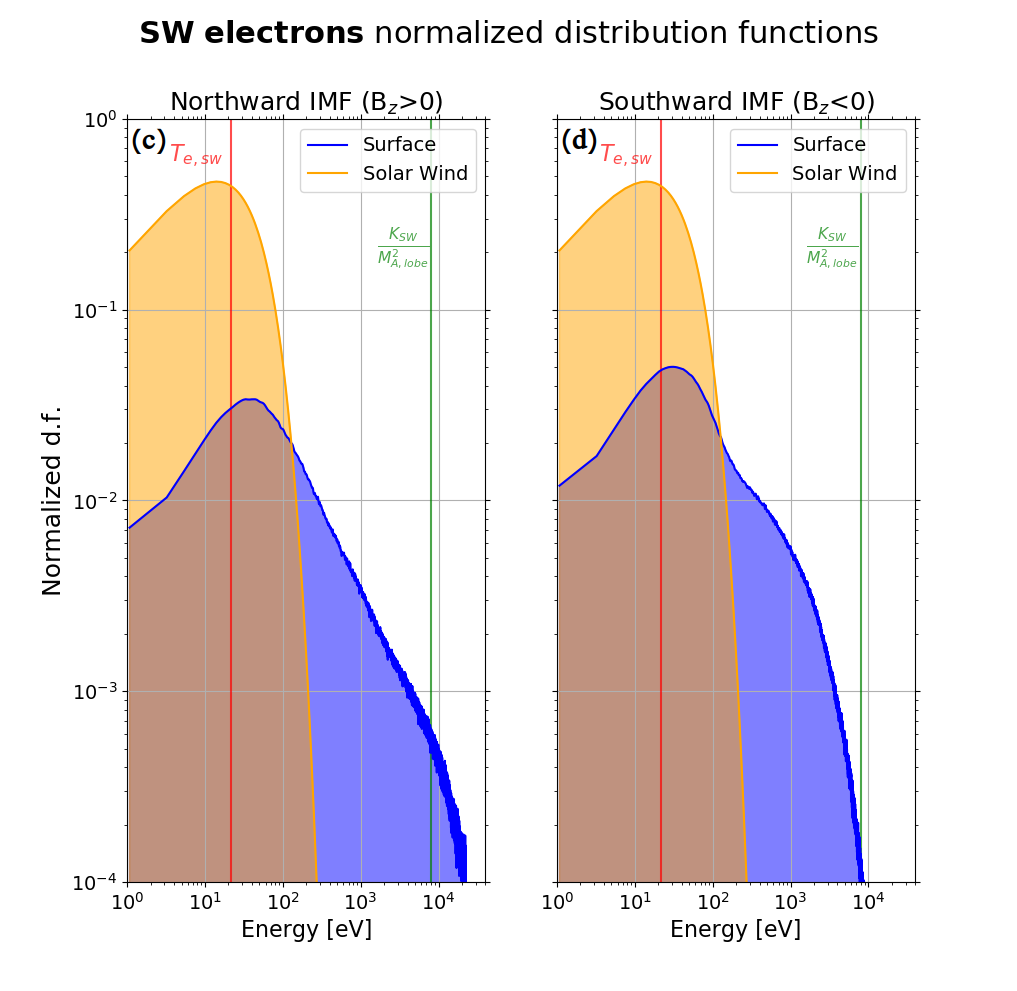}
    \caption{Normalized energy distribution of the particles in the solar wind (orange) and at the surface (blue).
    Panels (a)-(b) show the proton distribution functions.
    Panels (c)-(d) show the electron distribution functions.    
    The red vertical solid lines show the mean energy of the solar wind plasma, which is equal to $K_{_{\text{SW}}}=m_{\text{i}} V^2_{_{\text{SW}}}/2 = 826$~eV for protons, and to $T_{{\text e,_{\text{SW}}}} = 21.5$~eV for electrons. 
    The green vertical dashed lines in panels (a)-(b) show an estimate of the proton energy downstream of the bow shock equal to $K_{_{\text{SW}}}/M^2_{\text{f},_{\text{SW}}}$, where $M_{\text{f},_{\text{SW}}}=3.3$ is the fast magnetosonic Mach number in the solar wind upstream of the bow shock.
    The green vertical solid lines show an estimate of the maximum magnetic energy available for magnetic reconnection equal to $K_{_{\text{SW}}}/M^2_{\text{A,lobe}}$, where $M_{\text{A,lobe}}=0.32$ is the Alfv\'en Mach number in the lobes computed using a magnetic field of $100$~nT and a density of $3$~cm$^{-3}$.}
    \label{fig:df_electrons+ions}
\end{figure}
\subsection{Energy distribution of particles at the surface}\label{subs:energy_distrib}
For protons, the net effect of Mercury's magnetic field is to broaden their energy distribution. In our simulations, solar wind protons are initialized with a narrow Maxwellian energy distribution centered around $m_{\text{i}} V^2_{_{\text{SW}}}/2=826$~eV, corresponding to a solar wind speed of $V_{_{\text{SW}}} = 400$~km~s$^{-1}$ (shown by the red vertical solid line in Fig.~\ref{fig:df_electrons+ions}(a)-(b)). The distribution width is of the order of $T_{\text{i},_{\text{SW}}}=21.5$~eV, as shown by the orange curves in Fig.~\ref{fig:df_electrons+ions}(a)-(b). At the surface, the energy distribution of protons spread out to a few keV, with a considerable population centered around $\sim$~200 eV, as shown by the blue curves in Fig.~\ref{fig:df_electrons+ions}(a)-(b). Two competing processes are at play to slow down some of the solar wind protons and accelerate others.
The slowdown of solar wind protons is due to the presence of a bow shock in front of the planet. Upon passing through the bow shock, protons are decelerated from the (upstream) solar wind speed of $V_{_{\text{SW}}} = 400$~km~s$^{-1}$ to the (downstream) fast magnetosonic speed $V_{\text{f}} = (V^2_{\text{A,i}}+V^2_{\text{s,i}})^{1/2}$~\citep{Belmont2019}.
The fast magnetosonic speed in the solar wind is $V_{\text{f}} = 120$~km~s$^{-1}$, corresponding to a kinetic energy of 76~eV (shown by the green vertical dashed line in Fig.~\ref{fig:df_electrons+ions}(a)-(b)).
In principle, $V_{\text{f}}$ should be computed using the density, temperature and magnetic field values in the magnetosheath, but given the uncertainty associated to these values, we use here as a lower limit the fast magnetosonic speed in the solar wind, where, $V_{\text{A,i}}=B_{_{\text{SW}}}/\sqrt{4\pi m_{\text{i}} n_{_{\text{SW}}}}$ is the Alfv\'en speed, $V_{\text{s,i}}=\sqrt{\gamma T_{\text{i},_{\text{SW}}}/m_{\text{i}}}$ is the ion sound speed, and $\gamma=2$ is the adiabatic index of the plasma. 
The acceleration of solar wind protons is due to magnetic reconnection in the magnetosphere. For a northward IMF, magnetic reconnection is weakly driven at the lobes at high latitudes. For a southward IMF, magnetic reconnection is more strongly driven both at the nose and in the tail of the magnetosphere. Magnetic reconnection converts part of the magnetic energy stored in the planetary magnetic field configuration, corresponding to roughly 10 keV (shown by the green vertical solid line in Fig.~\ref{fig:df_electrons+ions}(a)-(b)), to kinetic energy of the particles. The efficiency of this conversion is on the order of 10\%~\citep{Shay2014,Phan2014,Haggerty2015}, meaning that roughly 1~keV of magnetic field energy is converted to kinetic energy of the particles.
These two competing processes (bow shock slowdown and acceleration by magnetic reconnection) are responsible for the proton fluences versus energy at the surface reported in Tab.~\ref{tab:tab1_results_maps}. Substantial proton fluences below 500~eV --mostly at high latitudes-- are a consequence of the plasma slowdown downstream of the bow shock. While, proton fluences above 1.5~keV --mostly at high latitudes (under northward IMF) or at low latitudes (under southward IMF)-- are a consequence of magnetic reconnection and proton heating in the magnetosheath. This result has implications for current estimates of ion sputtering at Mercury that consider ions with a fixed energy of 1~keV~amu$^{-1}$ (corresponding to the solar wind proton energy upstream of the bow shock). A broader discussion on the impact of proton energy distributions on sputtering yield can be found in Sec.~\ref{subs:disc_acceler}. 

For electrons, the net effect of Mercury's magnetic field is energization by magnetic reconnection. Solar wind electrons have a Maxwellian energy distribution with temperature $T_{\text{e},_{\text{SW}}}$=21.5~eV, as shown by the orange curves in Fig.~\ref{fig:df_electrons+ions}(c)-(d). For electrons, the flow energy component $m_{\text{e}} V^2_{_{\text{SW}}}/2$=8.26~eV (in the solar wind) is negligible with respect to their thermal energy component, both outside and inside Mercury's magnetosphere. At the surface, the electron energy distribution displays a high-energy tail extending up to $\sim$~10~keV, as shown by the blue curves in Fig.~\ref{fig:df_electrons+ions}(c)-(d). 
Magnetic reconnection energizes a substantial fraction of solar wind electrons up to the maximum magnetic energy available, shown by the green vertical solid line in Fig.~\ref{fig:df_electrons+ions}(c)-(d). 
As a consequence, precipitating electrons are mostly found in the energy range 100-4000 eV, as summarized by the fluences in Tab.~\ref{tab:tab1_results_maps}.

\section{Discussion and implications}\label{sec:discussion}

The study of solar wind particle precipitation on weakly magnetized bodies (such as Mercury) is more complex than that on unmagnetized bodies (such as the Moon). The presence of an intrinsic magnetic field -- although weak -- has a considerable effect on both \textit{where} particles precipitate onto the surface and \textit{what energy} these particles have. In the following, we discuss these two effects separately and their implications for plasma-surface interaction processes. Particular relevance is given to the comparison between weakly magnetized and unmagnetized bodies.

\subsection{Magnetosphere as a filter: the effect of a magnetic field on the spatial distribution of particles at the surface}\label{subs:disc_filter}

We consider here the interaction between an homogeneous, constant flow of plasma (i.e., an ideal solar wind) and a spherical body (i.e., an ideal rocky body such as Mercury or the Moon). The body rotates with angular velocity $\omega$ and spin axis perpendicular to the plasma flow.
If the body is unmagnetized, the flux of particles reaching the surface integrated over a time $\Delta T \gg 2\pi/\omega$ corresponds to the solar wind fluence multiplied by the cross-sectional area $\pi R^2$ and divided by the total surface body area $4\pi R^2$: 
\begin{equation}\label{eq:fluence_1}
    \frac{n_{_{\text{SW}}} V_{_{\text{SW}}} }{4} \Delta T
\end{equation}.

If the body is weakly magnetized, the fluence of particles at the surface is further reduced by a factor $\alpha > 1$ as follows:
\begin{equation}\label{eq:fluence_2}
    \frac{n_{_{\text{SW}}} V_{_{\text{SW}}} }{4 \alpha} \Delta T
\end{equation}.
The effective shielding parameter $\alpha$ accounts for the average reduction of plasma flux from the solar wind onto the surface due to the magnetic field. From the averaged fluences reported in Fig.~\ref{fig:fig2_means_i} for protons and Fig.~\ref{fig:fig2_means_e} for electrons, we derive the effective shielding parameters for solar wind protons and electrons:
\begin{eqnarray}\label{eq:alpha}
    \alpha_p &\approx& 20,\\
    \alpha_e &\approx& 10.
\end{eqnarray}
Our results indicate that the magnetic field effectively shields about 90\% of the incoming solar wind particles, reducing the fluence at the surface by one order of magnitude as compared to the unmagnetized case. The shielding values do not vary appreciably between our two runs using a purely northward or a purely southward IMF, suggesting that the IMF direction weakly affects the total number of particles precipitating onto the surface. A future study using a more realistic IMF direction for Mercury (including $B_x$ and $B_y$ components) will address this point specifically. 

The IMF direction controls \textit{where} solar wind particles precipitate onto the surface. Although the total number of precipitating particles weakly depends on the IMF direction, at least for the configurations under study, the regions of strongest precipitation significantly differs between the simulations with a purely northward and a purely southward IMF, as shown in Sec.~\ref{subs:spatial_distrib}.
Therefore, time variations in the IMF direction -- not included in our computation -- would modify the location of the regions of strongest particle precipitation. Since the IMF direction changes on time scales of the order of tens of minutes due to the turbulence in the solar wind~\citep{James2017}, we expect a continuously changing pattern of precipitation.
Therefore, the use of the effective shielding parameter $\alpha$ in Eq.~\ref{eq:fluence_2} represents a first order measure of a non-homogeneous and time-varying process. 

As recently highlighted by MESSENGER/FIPS observations of proton precipitation at the northern cusp~\citep{Raines2022}, one pervasive characteristic of Mercury's interaction with the solar wind is high variability. 
This study, however, focuses on representative cases with fixed solar wind conditions; we then computed the effect of these fixed cases over two full Mercury orbits. In doing so, several types of events which may provide substantial contributions to space weathering (such as coronal mass ejections (CMEs), interplanetary shocks and solar energetic particle (SEP) events) are omitted. These extreme events profoundly alter the ability of the planetary magnetic field to shield the surface. From MESSENGER observations,~\citet{Winslow2015} identified a total of 61 CMEs from March 2011 to September 2014, corresponding to an occurring frequency of roughly one CME every 20 days. The increase in ram pressure associated to these events compresses the dayside magnetopause, increases the surface exposed to solar wind flux and, in few abnormally strong and rare cases, rips off completely the dayside magnetosphere~\citep{Slavin2014,Winslow2017,Exner2018,Slavin2019,Jia2019}.
Concerning SEP events,~\citet{Gershman2015b} inferred the high-latitude precipitation of solar energetic electrons onto the surface of Mercury for 11 SEP events. 
As SEP events include protons over 15~MeV and electrons up to 3~MeV in energy at substantial fluxes, they can make substantial contributions to space weathering over periods of hours to days~\citep{Lario2013}. More details concerning the probability of all of these events occurring as well as their full impact on space weathering processes, however, remains poorly known at present.

Ion precipitation onto the surface of Mercury is one of the drivers of space weathering via ion sputtering and ion implantation. Space weathering alters the spectral properties of the regolith by darkening the surface, decreasing absorption band depths, and changing the spectral slope (reddening from the visible to near-infrared, bluing from the ultraviolet to the visible)~\citep{Noble2007,Blewett2021}. The infrared spectrometer MERTIS onboard BepiColombo will be able to measure those spectral variations at Mercury~\citep{Hiesinger2020,Maturilli2014}.
The geographical distributions of proton fluxes at the surface of Mercury shown in Fig.~\ref{fig:fig1_ions} when compared to spectral surface properties, can provide key information to infer the role of solar wind ion irradiation on the spectral processing of the surface. However, given the strong dependence of these maps on the time-variable IMF direction, we introduced the proton effective shielding parameter $\alpha_p \approx 20$ in Eq.~\ref{eq:fluence_2}. The use of this parameter will enable researchers to compare the partially shielded surface of Mercury with those unshielded of the Moon and of unmagnetized asteroids such as asteroids Ryugu~\citep{Hirotaka2017} and Bennu~\citep{Lauretta2017}, paving the way to comparative space weathering studies.

Space weathering studies of the hermean regolith rely on comparisons with the Moon and with asteroids~\citep{Domingue2014}.
Understanding how the magnetic field of Mercury shields the planetary surface compared to that of other unmagnetized bodies is a key point to study comparatively the effects of space weathering on solar system bodies. At Mercury, in this work, we showed that the surface is exposed on average to $\sim 2 \times 10^{14}$~protons~cm$^{-2}$ integrating over two full Mercury orbits. At the Moon, using Eq.~\ref{eq:fluence_1} and considering a solar wind flux of $3 \times 10^8$~protons~cm$^{-2}$s$^{-1}$, we find a fluence of $\sim 10^{15}$~protons~cm$^{-2}$ for the same time period (this value would be reduced to $\sim 7 \times 10^{14}$~protons~cm$^{-2}$ if considering zero proton flux when the Moon crosses Earth's magnetosphere;~\citet{Poppe2018}). For main belt asteroids, this fluence is further reduced by a factor $\sim 5 - 10$ due to the increased distance from the Sun. Therefore, Mercury is bombarded by roughly $3 - 5$ times fewer protons as compared to the Moon and by roughly the same amount of protons as compared to a main belt asteroid. At Mercury, the magnetic field screening compensates for the increase in solar wind flux, as compared to the Moon. 
In this work, we neglected the interplay between ion irradiation and micro-meteoroids processes such as impact gardening and comminution. Given that the micro-meteoroid impactor flux at Mercury is about a factor 5.5 higher than at the Moon~\citep{Cintala1992}, this interplay might be important and should be addressed in future works.

Space weathering of Mercury’s surface has consequences for the remotely sensed properties of the regolith, namely the determination of composition via color and spectral properties. Composition, both the identification of specific minerals and their abundances, are based on the detection and strength of absorption features. Space weathering, by both solar wind irradiation and micrometeoroid impacts, reduces the strength of absorption features. For example, the 1~$\mu$m band diagnostic of the mineral olivine can be completely masked by the formation of a small amount of nanophase iron ($\sim 1\%$ in weight of np-Fe$^0$), via solar wind irradiation and micro-meteoroid-impact-generated vapor~\citep{Kohout2014}. At Mercury, the MESSENGER infrared spectrometer MASCS/VNIR~\citep{McClintock2007} showed no distinct spectral features throughout the visible to near-infrared, except for the possible indication of sulfide mineralogy within the hollows~\citep{Vilas2016}. Given the reduced proton fluence onto Mercury’s surface compared to the Moon (by roughly a factor $3 - 5$), we suggest that ion irradiation is not the main process at play in reducing spectral band signatures at Mercury.

\subsection{Magnetosphere as an accelerator and decelerator: the effect of a magnetic field on the energy distribution of particles at the surface}\label{subs:disc_acceler}

Our numerical simulations show that the energy distribution of both protons and electrons are affected differently by the hermean magnetosphere. On the one hand, the protons impacting the surface are composed of one low-energy population (around $\sim$~200~eV) and of one moderate-to-high-energy population (around $\sim$~1~keV). On the other hand, the electrons impacting the surface have a moderate energy of the order of $\sim$~0.1-1~keV. The fluences of these different populations at the surface are summarized in Tab.~\ref{tab:tab1_results_maps}.

Our results on the proton energy distribution at the surface are key to reliably modeling the exosphere of Mercury. Solar wind protons contribute to the exosphere of Mercury via ion sputtering.
Ion sputtering is usually included in exosphere models assuming a monochromatic energy of 1~keV~amu$^{-1}$. This is a good approximation for protons in the solar wind, as shown by the orange curves in Fig.~\ref{fig:df_electrons+ions}(a)-(b). However, this is not a good approximation at the hermean surface where the proton distribution (in blue in Fig.~\ref{fig:df_electrons+ions}(a)-(b)) shows a large population at lower energies, coming from the interaction with the bow shock. Protons below $\sim$~500~eV weakly contribute to ion sputtering~\citep{Eckstein2007}. Our modeling suggests that only $\sim$~60\% of the total precipitating protons will  have significant sputtering yields. From the fluences in Tab.~\ref{tab:tab1_results_maps}, we see that this value changes with latitude. Around the poles, $\sim$~50\% of the protons are efficient for sputtering (fluence of $\sim 1.7 \times 10^{14}$~protons~cm$^{-2}$); whereas around the equator, this value goes to 100\% with a fluence of $\sim 0.7 \times 10^{14}$ protons~cm$^{-2}$. This mixture of spatial and energy dependence of the precipitating proton flux is a key ingredient that should be account for in the modeling of the Hermean magnetosphere and exosphere in the future.

Another major release mechanism for volatiles into Mercury's exosphere is photon stimulated desorption (PSD)~\citep{McGrath1986,Killen1990,Madey1998,Leblanc2022}; a process that is enhanced by ion bombardment~\citep{McGrath1986,Mura2009,Leblanc2022}. The coupling of these two processes is thought to produce a sodium exosphere in regions typical of ion sputtering, but with the high efficiency and lower energy distribution of PSD. Estimates and patterns of volatile release via PSD will be modulated by the actual flux and energies to the surface.

Solar wind electrons contribute to the exosphere of Mercury via electron stimulated desorption (ESD). This process is usually neglected in state-of-the-art exosphere models due to the lack of quantitative information on the flux and energy of precipitating electrons. 
Past works~\citep{McLain2011,Schriver2011b} estimated that ESD can generate as many neutrals as ion sputtering at Mercury; however, such estimates come with large uncertainties.
In our work, we provided quantitative estimates of the electron fluence and energy distribution at the surface of Mercury that will help to better evaluate the relevance of ESD as a source process for the exosphere of Mercury. The high-resolution maps in Fig.~\ref{fig:fig1_electrons} or the coarse grid values in Tab.~\ref{tab:tab1_results_maps} as well as the energy distribution in Fig.~\ref{fig:df_electrons+ions}(c)-(d) can be used to advance the exosphere modeling at Mercury including ESD.

Solar wind electrons in the range 0.5-10 keV drive X-rays emissions from the surface via X-ray fluorescence (XRF). While this process does not affect the chemistry of the surface, it allows the detection and mapping of regions of electron precipitation. Such emissions have been observed by the XRS instrument onboard MESSENGER for lines of Si K$\alpha$ (around 2 keV) and Ca K$\alpha$ (around 4 keV)~\citep{Lindsay2016,Lindsay2022}. These emissions are driven by keV electrons accelerated in the magnetosphere by magnetic reconnection, as shown by~\citet{Lavorenti2023} and discussed in Sec.~\ref{subs:energy_distrib}. However, XRS observations had a limited energy resolution and the surface coverage was constrained by MESSENGER's orbit. The MIXS instrument onboard BepiColombo~\citep{Bunce2020} will extend XRS observations by providing (i) more coverage in the southern hemisphere of Mercury, (ii) a higher energy resolution, allowing the separation of the Mg, Al, and Si lines, and (iii) a larger energy range enabling the observations of lines from Na, Fe and O. The novel capabilities of MIXS to detect more and lower energy fluorescence lines, coupled with electron observations from the Mio/MEA instrument~\citep{Saito2021}, will help to constraint the energy spectrum and source process of precipitating $\sim$keV electrons at Mercury. The electron precipitation maps in Fig.~\ref{fig:fig1_electrons} coupled with surface composition models will help to interpret the future MIXS observations of electron-induced X-ray fluorescence at Mercury. From the maps in Fig.~\ref{fig:fig1_electrons}, indirect information on the IMF direction can also be derived from the distribution of X-ray fluorescence onto the surface.

Surface charging at Mercury remains an open question at present. From the precipitating fluxes of protons and electrons computed in Sec.~\ref{sec:results}, we note a clear tendency of the surface to be negatively charged (the electrons mean fluence is around a factor 2 higher than that of protons, as shown in Tab.~\ref{tab:tab1_results_maps}), at least on the nightside of the planet or within shadow in the dayside. This is somewhat expected given the higher mobility of electrons as compared to protons. Nonetheless, at this stage, it is hard to make any definitive conclusions about surface charging from our simulations given (i) the lack of sub-Debye-length kinetic physics in our implicit PIC code, and (ii) the lack of photo-electron emission from the surface of Mercury in our model that is essential to constraining surface charging in illuminated regions of the hermean surface. Future works could address this question using explicit PIC codes resolving the sub-Debye-length plasma-surface interactions as well as including photo-electron emission from the illuminated surface. Protons and electrons fluxes from the maps in this work (in Figs.~\ref{fig:fig1_ions}-\ref{fig:fig2_means_e}) could then be used as boundary conditions for such smaller scale simulations.

\section{Conclusion}\label{sec:conclusion}
From numerical simulations of the interaction of Mercury's magnetic field with solar wind plasma (protons and electrons), we have computed particle precipitation maps at the planetary surface integrated over two full Mercury orbits (176 Earth days). Our results show that:
\begin{itemize}
    \item Mercury's 3:2 spin-orbit resonance has a weak effect on the time-integrated plasma precipitation pattern onto the surface. The surface pattern of precipitating plasma more strongly depends on the upstream magnetic field direction.
    \item Mercury's weak magnetic field is able to shield on average 90\% of impinging solar wind protons and electrons. 
    \item Mercury's bow shock and magnetosphere tend to broaden the proton energy distribution, from the solar wind onto the surface. A considerable number of protons at the surface have energies below 500~eV (mostly at high latitudes) and above 1.5~keV (mostly at low latitudes).
    \item Mercury's magnetosphere tend to extend to high energies the electron energy distribution, with most of the electrons at the surface found in the range 0.1-4~keV.
\end{itemize}
Our results demonstrate the complexity of Mercury's geographical plasma precipitation and paves the way for future quantitative studies addressing (i) space weathering of Mercury's regolith, (ii) plasma-driven source processes for Mercury's exosphere such as ion sputtering and ESD, and (iii) electron-induced X-ray fluorescent emission from Mercury's regolith.
The data used throughout the paper are publicly available at~\citet{Jensen2023_zenodo}.

\section{Acknowledgments}
FL and PH acknowledges TGCC under the allocations AP010412622 and A0100412428. FL acknowledges the CINECA award under the ISCRA initiative, for the availability of high performance computing resources and support. FL and PH acknowledge the support of CNES for the BepiColombo mission. FL acknowledges ESA support of the PhD. This research was supported by the ISSI in Bern (ISSI International Team project \#525). JR was supported by NASA Discovery Data Analysis grant 80NSSC20K1148. SA is supported by JSPS KAKENHI number: 22J01606. DD, EAJ, and DWS are supported by NASA Solar System Workings Grant 80NSSC22K0099.

\bibliography{biblio}{}

\begin{thebibliography}{}
\expandafter\ifx\csname natexlab\endcsname\relax\def\natexlab#1{#1}\fi
\providecommand{\url}[1]{\href{#1}{#1}}
\providecommand{\dodoi}[1]{doi:~\href{http://doi.org/#1}{\nolinkurl{#1}}}
\providecommand{\doeprint}[1]{\href{http://ascl.net/#1}{\nolinkurl{http://ascl.net/#1}}}
\providecommand{\doarXiv}[1]{\href{https://arxiv.org/abs/#1}{\nolinkurl{https://arxiv.org/abs/#1}}}

\bibitem[{Bauch {et~al.}(2021)Bauch, Hiesinger, Greenhagen, \&
  Helbert}]{Bauch2021}
Bauch, K.~E., Hiesinger, H., Greenhagen, B.~T., \& Helbert, J. 2021, Icarus,
  354, 114083, \dodoi{10.1016/j.icarus.2020.114083}

\bibitem[{Belmont {et~al.}(2019)Belmont, Rezeau, Riconda, \&
  Zaslavsky}]{Belmont2019}
Belmont, G., Rezeau, L., Riconda, C., \& Zaslavsky, A. 2019, in Introduction to
  Plasma Physics, ed. G.~Belmont, L.~Rezeau, C.~Riconda, \& A.~Zaslavsky
  (Elsevier), 195--218, \dodoi{10.1016/B978-1-78548-306-6.50007-X}

\bibitem[{{Benkhoff} {et~al.}(2021){Benkhoff}, {Murakami}, {Baumjohann},
  {Besse}, {Bunce}, {Casale}, {Cremosese}, {Glassmeier}, {Hayakawa}, {Heyner},
  {Hiesinger}, {Huovelin}, {Hussmann}, {Iafolla}, {Iess}, {Kasaba},
  {Kobayashi}, {Milillo}, {Mitrofanov}, {Montagnon}, {Novara}, {Orsini},
  {Quemerais}, {Reininghaus}, {Saito}, {Santoli}, {Stramaccioni}, {Sutherland},
  {Thomas}, {Yoshikawa}, \& {Zender}}]{Benkhoff2021}
{Benkhoff}, J., {Murakami}, G., {Baumjohann}, W., {et~al.} 2021, Space Science
  Reviews, 217, 90, \dodoi{10.1007/s11214-021-00861-4}

\bibitem[{{Benna} {et~al.}(2010){Benna}, {Anderson}, {Baker}, {Boardsen},
  {Gloeckler}, {Gold}, {Ho}, {Killen}, {Korth}, {Krimigis}, {Purucker},
  {McNutt}, {Raines}, {McClintock}, {Sarantos}, {Slavin}, {Solomon}, \&
  {Zurbuchen}}]{Benna2010}
{Benna}, M., {Anderson}, B.~J., {Baker}, D.~N., {et~al.} 2010, Icarus, 209, 3,
  \dodoi{10.1016/j.icarus.2009.11.036}

\bibitem[{{Blewett} {et~al.}(2021){Blewett}, {Denevi}, {Cahill}, \&
  {Klima}}]{Blewett2021}
{Blewett}, D.~T., {Denevi}, B.~W., {Cahill}, J. T.~S., \& {Klima}, R.~L. 2021,
  Icarus, 364, 114472, \dodoi{10.1016/j.icarus.2021.114472}

\bibitem[{{Bunce} {et~al.}(2020){Bunce}, {Martindale}, {Lindsay}, {Muinonen},
  {Rothery}, {Pearson}, {McDonnell}, {Thomas}, {Thornhill}, {Tikkanen},
  {Feldman}, {Huovelin}, {Korpela}, {Esko}, {Lehtolainen}, {Treis}, {Majewski},
  {Hilchenbach}, {V{\"a}is{\"a}nen}, {Luttinen}, {Kohout}, {Penttil{\"a}},
  {Bridges}, {Joy}, {Alcacera-Gil}, {Alibert}, {Anand}, {Bannister},
  {Barcelo-Garcia}, {Bicknell}, {Blake}, {Bland}, {Butcher}, {Cheney},
  {Christensen}, {Crawford}, {Crawford}, {Dennerl}, {Dougherty}, {Drumm},
  {Fairbend}, {Genzer}, {Grande}, {Hall}, {Hodnett}, {Houghton}, {Imber},
  {Kallio}, {Lara}, {Balado Margeli}, {Mas-Hesse}, {Maurice}, {Milan},
  {Millington-Hotze}, {Nenonen}, {Nittler}, {Okada}, {Orm{\"o}},
  {Perez-Mercader}, {Poyner}, {Robert}, {Ross}, {Pajas-Sanz}, {Schyns},
  {Seguy}, {Str{\"u}der}, {Vaudon}, {Viceira-Mart{\'\i}n}, {Williams},
  {Willingale}, \& {Yeoman}}]{Bunce2020}
{Bunce}, E.~J., {Martindale}, A., {Lindsay}, S., {et~al.} 2020, Space Science
  Reviews, 216, 126, \dodoi{10.1007/s11214-020-00750-2}

\bibitem[{{Cintala}(1992)}]{Cintala1992}
{Cintala}, M.~J. 1992, \jgr, 97, 947, \dodoi{10.1029/91JE02207}

\bibitem[{{Domingue} {et~al.}(2014){Domingue}, {Chapman}, {Killen},
  {Zurbuchen}, {Gilbert}, {Sarantos}, {Benna}, {Slavin}, {Schriver},
  {Tr{\'a}vn{\'\i}{\v{c}}ek}, {Orlando}, {Sprague}, {Blewett}, {Gillis-Davis},
  {Feldman}, {Lawrence}, {Ho}, {Ebel}, {Nittler}, {Vilas}, {Pieters},
  {Solomon}, {Johnson}, {Winslow}, {Helbert}, {Peplowski}, {Weider}, {Mouawad},
  {Izenberg}, \& {McClintock}}]{Domingue2014}
{Domingue}, D.~L., {Chapman}, C.~R., {Killen}, R.~M., {et~al.} 2014, Space
  Science Reviews, 181, 121, \dodoi{10.1007/s11214-014-0039-5}

\bibitem[{Eckstein(2007)}]{Eckstein2007}
Eckstein, W. 2007, Sputtering Yields (Berlin, Heidelberg: Springer Berlin
  Heidelberg), 33--187, \dodoi{10.1007/978-3-540-44502-9_3}

\bibitem[{Exner {et~al.}(2018)Exner, Heyner, Liuzzo, Motschmann, Shiota,
  Kusano, \& Shibayama}]{Exner2018}
Exner, W., Heyner, D., Liuzzo, L., {et~al.} 2018, Planetary and Space Science,
  153, 89, \dodoi{10.1016/j.pss.2017.12.016}

\bibitem[{{Fatemi} {et~al.}(2020){Fatemi}, {Poppe}, \& {Barabash}}]{Fatemi2020}
{Fatemi}, S., {Poppe}, A.~R., \& {Barabash}, S. 2020, Journal of Geophysical
  Research (Space Physics), 125, e27706, \dodoi{10.1029/2019JA027706}

\bibitem[{{Gershman} {et~al.}(2015){Gershman}, {Raines}, {Slavin}, {Zurbuchen},
  {Anderson}, {Korth}, {Ho}, {Boardsen}, {Cassidy}, {Walsh}, \&
  {Solomon}}]{Gershman2015b}
{Gershman}, D.~J., {Raines}, J.~M., {Slavin}, J.~A., {et~al.} 2015, Journal of
  Geophysical Research (Space Physics), 120, 8559, \dodoi{10.1002/2015JA021610}

\bibitem[{{Glass} {et~al.}(2022){Glass}, {Raines}, {Jia}, {Sun}, {Imber},
  {Dewey}, \& {Slavin}}]{Glass2022}
{Glass}, A.~N., {Raines}, J.~M., {Jia}, X., {et~al.} 2022, Journal of
  Geophysical Research (Space Physics), 127, e2022JA030969,
  \dodoi{10.1029/2022JA030969}

\bibitem[{{Haggerty} {et~al.}(2015){Haggerty}, {Shay}, {Drake}, {Phan}, \&
  {McHugh}}]{Haggerty2015}
{Haggerty}, C.~C., {Shay}, M.~A., {Drake}, J.~F., {Phan}, T.~D., \& {McHugh},
  C.~T. 2015, Geophysics Research Letters, 42, 9657,
  \dodoi{10.1002/2015GL065961}

\bibitem[{{Hiesinger} {et~al.}(2020){Hiesinger}, {Helbert}, {Alemanno},
  {Bauch}, {D'Amore}, {Maturilli}, {Morlok}, {Reitze}, {Stangarone}, {Stojic},
  {Varatharajan}, {Weber}, \& {Mertis Co-I Team}}]{Hiesinger2020}
{Hiesinger}, H., {Helbert}, J., {Alemanno}, G., {et~al.} 2020, Space Science
  Reviews, 216, 110, \dodoi{10.1007/s11214-020-00732-4}

\bibitem[{James {et~al.}(2017)James, Imber, Bunce, Yeoman, Lockwood, Owens, \&
  Slavin}]{James2017}
James, M.~K., Imber, S.~M., Bunce, E.~J., {et~al.} 2017, Journal of Geophysical
  Research: Space Physics, 122, 7907, \dodoi{10.1002/2017JA024435}

\bibitem[{{Jia} {et~al.}(2019){Jia}, {Slavin}, {Poh}, {DiBraccio}, {Toth},
  {Chen}, {Raines}, \& {Gombosi}}]{Jia2019}
{Jia}, X., {Slavin}, J.~A., {Poh}, G., {et~al.} 2019, Journal of Geophysical
  Research (Space Physics), 124, 229, \dodoi{10.1029/2018JA026166}

\bibitem[{{Killen} {et~al.}(2022){Killen}, {Morrissey}, {Burger}, {Vervack},
  {Tucker}, \& {Savin}}]{Killen2022}
{Killen}, R.~M., {Morrissey}, L.~S., {Burger}, M.~H., {et~al.} 2022, \psj, 3,
  139, \dodoi{10.3847/PSJ/ac67de}

\bibitem[{{Killen} {et~al.}(1990){Killen}, {Potter}, \& {Morgan}}]{Killen1990}
{Killen}, R.~M., {Potter}, A.~E., \& {Morgan}, T.~H. 1990, \icarus, 85, 145,
  \dodoi{10.1016/0019-1035(90)90108-L}

\bibitem[{Kohout {et~al.}(2014)Kohout, {\v{C}}uda, Filip, Britt, Bradley,
  Tu{\v{c}}ek, Sk{\'a}la, Kletetschka, Ka{\v{s}}l{\'\i}k, Malina,
  {et~al.}}]{Kohout2014}
Kohout, T., {\v{C}}uda, J., Filip, J., {et~al.} 2014, Icarus, 237, 75

\bibitem[{Lario {et~al.}(2013)Lario, Aran, Gómez-Herrero, Dresing, Heber, Ho,
  Decker, \& Roelof}]{Lario2013}
Lario, D., Aran, A., Gómez-Herrero, R., {et~al.} 2013, The Astrophysical
  Journal, 767, 41, \dodoi{10.1088/0004-637X/767/1/41}

\bibitem[{{Lauretta} {et~al.}(2017){Lauretta}, {Balram-Knutson}, {Beshore},
  {Boynton}, {Drouet d'Aubigny}, {DellaGiustina}, {Enos}, {Golish},
  {Hergenrother}, {Howell}, {Bennett}, {Morton}, {Nolan}, {Rizk}, {Roper},
  {Bartels}, {Bos}, {Dworkin}, {Highsmith}, {Lorenz}, {Lim}, {Mink}, {Moreau},
  {Nuth}, {Reuter}, {Simon}, {Bierhaus}, {Bryan}, {Ballouz}, {Barnouin},
  {Binzel}, {Bottke}, {Hamilton}, {Walsh}, {Chesley}, {Christensen}, {Clark},
  {Connolly}, {Crombie}, {Daly}, {Emery}, {McCoy}, {McMahon}, {Scheeres},
  {Messenger}, {Nakamura-Messenger}, {Righter}, \& {Sandford}}]{Lauretta2017}
{Lauretta}, D.~S., {Balram-Knutson}, S.~S., {Beshore}, E., {et~al.} 2017, \ssr,
  212, 925, \dodoi{10.1007/s11214-017-0405-1}

\bibitem[{{Lavorenti} {et~al.}(2022){Lavorenti}, {Henri}, {Califano}, {Deca},
  {Aizawa}, {Andr{\'e}}, \& {Benkhoff}}]{Lavorenti2022}
{Lavorenti}, F., {Henri}, P., {Califano}, F., {et~al.} 2022, \aap, 664, A133,
  \dodoi{10.1051/0004-6361/202243911}

\bibitem[{{Lavorenti} {et~al.}(2023){Lavorenti}, {Henri}, {Califano}, {Deca},
  {Lindsay}, {Aizawa}, \& {Benkhoff}}]{Lavorenti2023}
---. 2023, \aap, accepted

\bibitem[{Lavorenti {et~al.}(2023)Lavorenti, Jensen, Aizawa, Califano, D'Amore,
  Domingue, Henri, Lindsay, Raines, \& Savin}]{Jensen2023_zenodo}
Lavorenti, F., Jensen, E., Aizawa, S., {et~al.} 2023, {Maps of solar wind
  plasma precipitation onto Mercury's surface: a geographical perspective},
  Zenodo, \dodoi{10.5281/zenodo.7927373}

\bibitem[{{Leblanc} {et~al.}(2022){Leblanc}, {Schmidt}, {Mangano}, {Mura},
  {Cremonese}, {Raines}, {Jasinski}, {Sarantos}, {Milillo}, {Killen},
  {Massetti}, {Cassidy}, {Vervack}, {Kameda}, {Capria}, {Horanyi}, {Janches},
  {Berezhnoy}, {Christou}, {Hirai}, {Lierle}, \& {Morgenthaler}}]{Leblanc2022}
{Leblanc}, F., {Schmidt}, C., {Mangano}, V., {et~al.} 2022, \ssr, 218, 2,
  \dodoi{10.1007/s11214-022-00871-w}

\bibitem[{{Lindsay} {et~al.}(2022){Lindsay}, {Bunce}, {Imber}, {Martindale},
  {Nittler}, \& {Yeoman}}]{Lindsay2022}
{Lindsay}, S.~T., {Bunce}, E.~J., {Imber}, S.~M., {et~al.} 2022, Journal of
  Geophysical Research (Space Physics), 127, e29675,
  \dodoi{10.1029/2021JA029675}

\bibitem[{{Lindsay} {et~al.}(2016){Lindsay}, {James}, {Bunce}, {Imber},
  {Korth}, {Martindale}, \& {Yeoman}}]{Lindsay2016}
{Lindsay}, S.~T., {James}, M.~K., {Bunce}, E.~J., {et~al.} 2016, Planetary and
  Space Science, 125, 72, \dodoi{10.1016/j.pss.2016.03.005}

\bibitem[{{Madey} {et~al.}(1998){Madey}, {Yakshinskiy}, {Ageev}, \&
  {Johnson}}]{Madey1998}
{Madey}, T.~E., {Yakshinskiy}, B.~V., {Ageev}, V.~N., \& {Johnson}, R.~E. 1998,
  Journal Geophysical Research, 103, 5873, \dodoi{10.1029/98JE00230}

\bibitem[{{Mangano} {et~al.}(2015){Mangano}, {Massetti}, {Milillo}, {Plainaki},
  {Orsini}, {Rispoli}, \& {Leblanc}}]{Mangano2015}
{Mangano}, V., {Massetti}, S., {Milillo}, A., {et~al.} 2015, \planss, 115, 102,
  \dodoi{10.1016/j.pss.2015.04.001}

\bibitem[{Markidis {et~al.}(2010)Markidis, Lapenta, \&
  Rizwan-uddin}]{Markidis2010}
Markidis, S., Lapenta, G., \& Rizwan-uddin. 2010, Mathematics and Computers in
  Simulation, 80, 1509, \dodoi{10.1016/j.matcom.2009.08.038}

\bibitem[{{Maturilli} {et~al.}(2014){Maturilli}, {Helbert}, {St. John}, {Head},
  {Vaughan}, {D'Amore}, {Gottschalk}, \& {Ferrari}}]{Maturilli2014}
{Maturilli}, A., {Helbert}, J., {St. John}, J.~M., {et~al.} 2014, Earth and
  Planetary Science Letters, 398, 58, \dodoi{10.1016/j.epsl.2014.04.035}

\bibitem[{{McClintock} \& {Lankton}(2007)}]{McClintock2007}
{McClintock}, W.~E., \& {Lankton}, M.~R. 2007, \ssr, 131, 481,
  \dodoi{10.1007/s11214-007-9264-5}

\bibitem[{{McGrath} {et~al.}(1986){McGrath}, {Johnson}, \&
  {Lanzerotti}}]{McGrath1986}
{McGrath}, M.~A., {Johnson}, R.~E., \& {Lanzerotti}, L.~J. 1986, \nat, 323,
  694, \dodoi{10.1038/323694a0}

\bibitem[{{McLain} {et~al.}(2011){McLain}, {Sprague}, {Grieves}, {Schriver},
  {Travinicek}, \& {Orlando}}]{McLain2011}
{McLain}, J.~L., {Sprague}, A.~L., {Grieves}, G.~A., {et~al.} 2011, Journal of
  Geophysical Research (Planets), 116, E03007, \dodoi{10.1029/2010JE003714}

\bibitem[{{Meyer-Vernet}(2007)}]{Meyer-Vernet2007}
{Meyer-Vernet}, N. 2007, {Basics of the Solar Wind}

\bibitem[{{Mura} {et~al.}(2009){Mura}, {Wurz}, {Lichtenegger}, {Schleicher},
  {Lammer}, {Delcourt}, {Milillo}, {Orsini}, {Massetti}, \&
  {Khodachenko}}]{Mura2009}
{Mura}, A., {Wurz}, P., {Lichtenegger}, H. I.~M., {et~al.} 2009, \icarus, 200,
  1, \dodoi{10.1016/j.icarus.2008.11.014}

\bibitem[{{Noble} {et~al.}(2007){Noble}, {Pieters}, \& {Keller}}]{Noble2007}
{Noble}, S.~K., {Pieters}, C.~M., \& {Keller}, L.~P. 2007, Icarus, 192, 629,
  \dodoi{10.1016/j.icarus.2007.07.021}

\bibitem[{{Ogilvie} {et~al.}(1974){Ogilvie}, {Scudder}, {Hartle}, {Siscoe},
  {Bridge}, {Lazarus}, {Asbridge}, {Bame}, \& {Yeates}}]{Ogilvie1974}
{Ogilvie}, K.~W., {Scudder}, J.~D., {Hartle}, R.~E., {et~al.} 1974, Science,
  185, 145, \dodoi{10.1126/science.185.4146.145}

\bibitem[{{Phan} {et~al.}(2014){Phan}, {Drake}, {Shay}, {Gosling}, {Paschmann},
  {Eastwood}, {Oieroset}, {Fujimoto}, \& {Angelopoulos}}]{Phan2014}
{Phan}, T.~D., {Drake}, J.~F., {Shay}, M.~A., {et~al.} 2014, Geophysics
  Research Letters, 41, 7002, \dodoi{10.1002/2014GL061547}

\bibitem[{{Poppe} {et~al.}(2018){Poppe}, {Farrell}, \& {Halekas}}]{Poppe2018}
{Poppe}, A.~R., {Farrell}, W.~M., \& {Halekas}, J.~S. 2018, Journal of
  Geophysical Research (Planets), 123, 37, \dodoi{10.1002/2017JE005426}

\bibitem[{{Raines} {et~al.}(2022){Raines}, {Dewey}, {Staudacher}, {Tracy},
  {Bert}, {Sarantos}, {Gershman}, {Jasinski}, {Bowers}, {Fisher}, \&
  {Slavin}}]{Raines2022}
{Raines}, J.~M., {Dewey}, R.~M., {Staudacher}, N.~M., {et~al.} 2022, Journal of
  Geophysical Research (Space Physics), 127, e2022JA030397,
  \dodoi{10.1029/2022JA030397}

\bibitem[{{Saito} {et~al.}(2021){Saito}, {Delcourt}, {Hirahara}, {Barabash},
  {Andr{\'e}}, {Takashima}, {Asamura}, {Yokota}, {Wieser}, {Nishino}, {Oka},
  {Futaana}, {Harada}, {Sauvaud}, {Louarn}, {Lavraud}, {G{\'e}not}, {Mazelle},
  {Dandouras}, {Jacquey}, {Aoustin}, {Barthe}, {Cadu}, {Fedorov}, {Frezoul},
  {Garat}, {Le Comte}, {Lee}, {M{\'e}dale}, {Moirin}, {Penou}, {Petiot},
  {Peyre}, {Rouzaud}, {S{\'e}ran}, {N{\"e}me{\v{c}}ek},
  {S{\c{t}}afr{\'a}nkov{\'a}}, {Marcucci}, {Bruno}, {Consolini}, {Miyake},
  {Shinohara}, {Hasegawa}, {Seki}, {Coates}, {Leblanc}, {Verdeil}, {Katra},
  {Fontaine}, {Illiano}, {Berthelier}, {Techer}, {Fraenz}, {Fischer}, {Krupp},
  {Woch}, {B{\"u}hrke}, {Fiethe}, {Michalik}, {Matsumoto}, {Yanagimachi},
  {Miyoshi}, {Mitani}, {Shimoyama}, {Zong}, {Wurz}, {Andersson}, {Karlsson},
  {Holmstr{\"o}m}, {Kazama}, {Ip}, {Hoshino}, {Fujimoto}, {Terada}, {Keika}, \&
  {BepiColombo Mio/MPPE Team}}]{Saito2021}
{Saito}, Y., {Delcourt}, D., {Hirahara}, M., {et~al.} 2021, Space Science
  Reviews, 217, 70, \dodoi{10.1007/s11214-021-00839-2}

\bibitem[{{Sarantos} {et~al.}(2007){Sarantos}, {Killen}, \&
  {Kim}}]{Sarantos2007}
{Sarantos}, M., {Killen}, R.~M., \& {Kim}, D. 2007, Planetary and Space
  Science, 55, 1584, \dodoi{10.1016/j.pss.2006.10.011}

\bibitem[{{Sawada} {et~al.}(2017){Sawada}, {Okazaki}, {Tachibana}, {Sakamoto},
  {Takano}, {Okamoto}, {Yano}, {Miura}, {Abe}, {Hasegawa}, \&
  {Noguchi}}]{Hirotaka2017}
{Sawada}, H., {Okazaki}, R., {Tachibana}, S., {et~al.} 2017, \ssr, 208, 81,
  \dodoi{10.1007/s11214-017-0338-8}

\bibitem[{{Schriver} {et~al.}(2011){Schriver}, {Tr{\'a}vn{\'\i}{\v{c}}ek},
  {Ashour-Abdalla}, {Richard}, {Hellinger}, {Slavin}, {Anderson}, {Baker},
  {Benna}, {Boardsen}, {Gold}, {Ho}, {Korth}, {Krimigis}, {McClintock},
  {McLain}, {Orlando}, {Sarantos}, {Sprague}, \& {Starr}}]{Schriver2011b}
{Schriver}, D., {Tr{\'a}vn{\'\i}{\v{c}}ek}, P., {Ashour-Abdalla}, M., {et~al.}
  2011, Planetary and Space Science, 59, 2026,
  \dodoi{10.1016/j.pss.2011.03.008}

\bibitem[{{Shay} {et~al.}(2014){Shay}, {Haggerty}, {Phan}, {Drake}, {Cassak},
  {Wu}, {Oieroset}, {Swisdak}, \& {Malakit}}]{Shay2014}
{Shay}, M.~A., {Haggerty}, C.~C., {Phan}, T.~D., {et~al.} 2014, Physics of
  Plasmas, 21, 122902, \dodoi{10.1063/1.4904203}

\bibitem[{Slavin {et~al.}(2014)Slavin, DiBraccio, Gershman, Imber, Poh, Raines,
  Zurbuchen, Jia, Baker, Glassmeier, Livi, Boardsen, Cassidy, Sarantos,
  Sundberg, Masters, Johnson, Winslow, Anderson, Korth, McNutt~Jr., \&
  Solomon}]{Slavin2014}
Slavin, J.~A., DiBraccio, G.~A., Gershman, D.~J., {et~al.} 2014, Journal of
  Geophysical Research: Space Physics, 119, 8087, \dodoi{10.1002/2014JA020319}

\bibitem[{{Slavin} {et~al.}(2019){Slavin}, {Middleton}, {Raines}, {Jia},
  {Zhong}, {Sun}, {Livi}, {Imber}, {Poh}, {Akhavan-Tafti}, {Jasinski},
  {DiBraccio}, {Dong}, {Dewey}, \& {Mays}}]{Slavin2019}
{Slavin}, J.~A., {Middleton}, H.~R., {Raines}, J.~M., {et~al.} 2019, Journal of
  Geophysical Research (Space Physics), 124, 6613, \dodoi{10.1029/2019JA026892}

\bibitem[{{Solomon} {et~al.}(2007){Solomon}, {McNutt}, {Gold}, \&
  {Domingue}}]{Solomon2007}
{Solomon}, S.~C., {McNutt}, R.~L., {Gold}, R.~E., \& {Domingue}, D.~L. 2007,
  Space Science Reviews, 131, 3, \dodoi{10.1007/s11214-007-9247-6}

\bibitem[{Vilas {et~al.}(2016)Vilas, Domingue, Helbert, D'Amore, Maturilli,
  Klima, {Stockstill-Cahill}, Murchie, Izenberg, Blewett, Vaughan, \&
  Head}]{Vilas2016}
Vilas, F., Domingue, D.~L., Helbert, J., {et~al.} 2016, Geophysical Research
  Letters, 43, 1450, \dodoi{10/gg3j2v}

\bibitem[{{Winslow} {et~al.}(2015){Winslow}, {Lugaz}, {Philpott}, {Schwadron},
  {Farrugia}, {Anderson}, \& {Smith}}]{Winslow2015}
{Winslow}, R.~M., {Lugaz}, N., {Philpott}, L.~C., {et~al.} 2015, Journal of
  Geophysical Research (Space Physics), 120, 6101, \dodoi{10.1002/2015JA021200}

\bibitem[{{Winslow} {et~al.}(2017){Winslow}, {Philpott}, {Paty}, {Lugaz},
  {Schwadron}, {Johnson}, \& {Korth}}]{Winslow2017}
{Winslow}, R.~M., {Philpott}, L., {Paty}, C.~S., {et~al.} 2017, Journal of
  Geophysical Research (Space Physics), 122, 4960, \dodoi{10.1002/2016JA023548}

\bibitem[{{Wurz} {et~al.}(2022){Wurz}, {Fatemi}, {Galli}, {Halekas}, {Harada},
  {J{\"a}ggi}, {Jasinski}, {Lammer}, {Lindsay}, {Nishino}, {Orlando}, {Raines},
  {Scherf}, {Slavin}, {Vorburger}, \& {Winslow}}]{Wurz2022}
{Wurz}, P., {Fatemi}, S., {Galli}, A., {et~al.} 2022, Space Science Reviews,
  218, 10, \dodoi{10.1007/s11214-022-00875-6}

\end{thebibliography}
\bibliographystyle{aasjournal}

\end{document}